\documentclass[prl,superscriptaddress,twocolumn]{revtex4}
\usepackage{enumerate}
\usepackage{amsfonts,amssymb,amsmath}
\usepackage[]{graphics,graphicx,epsfig}
\usepackage{amsthm}

\usepackage{marvosym}

\usepackage{fdsymbol}
\usepackage{graphicx}
\usepackage{dcolumn}
\usepackage{natbib}
\usepackage{color}
\usepackage{xcolor}
\usepackage{multirow}
\usepackage{ulem}
\newcommand{\ket}[1]{|{#1}\rangle}
\newcommand{\bra}[1]{\langle{#1}|}

\newcommand{\braket}[2]{\langle{#1}|{#2}\rangle}
\newcommand\myprime{\mkern-3.5mu\raise0.8ex\hbox{$\scriptstyle\prime$}}
\begin{document}


\title{Exposing Hypersensitivity in Quantum Chaotic Dynamics}

\author{Andrzej Grudka}
\affiliation{Institute of Spintronics and Quantum Information, Faculty of Physics, Adam Mickiewicz University, 61-614 Pozna\'n, Poland}

\author{Pawe{\l} Kurzy{\'n}ski}
\email{pawel.kurzynski@amu.edu.pl}
\affiliation{Institute of Spintronics and Quantum Information, Faculty of Physics, Adam Mickiewicz University, 61-614 Pozna\'n, Poland}

\author{Adam S. Sajna}
\affiliation{Institute of Theoretical Physics, Faculty of Fundamental Problems of Technology, Wroclaw University of Science and Technology, 50-370 Wroclaw, Poland}

\author{Jan W{\'o}jcik}
\affiliation{Faculty of Physics, Adam Mickiewicz University, 61-614 Pozna\'n, Poland}

\author{Antoni W{\'o}jcik}
\affiliation{Institute of Spintronics and Quantum Information, Faculty of Physics, Adam Mickiewicz University, 61-614 Pozna\'n, Poland}

\date{\today}

\newcommand{\adam}{\color{red}}


\begin{abstract}
We demonstrate that the unitary dynamics of a multi-qubit system can display hypersensitivity to initial state perturbation. This contradicts the common belief that the classical approach based on the exponential divergence of initially neighboring trajectories cannot be applied to identify chaos in quantum systems. To observe hypersensitivity we use quantum state-metric, introduced by Girolami and Anza in [Phys. Rev. Lett. 126 (2021) 170502], which can be interpreted as a quantum Hamming distance. As an example of a quantum system, we take the multi-qubit implementation of the quantum kicked top, a paradigmatic system known to exhibit quantum chaotic behavior. Our findings confirm that the observed hypersensitivity corresponds to commonly used signatures of quantum chaos. Furthermore, we demonstrate that the proposed metric can detect quantum chaos in the same regime and under analogous initial conditions as in the corresponding classical case.
\end{abstract}

\maketitle


{\it Introduction.} Hypersensitivity to small perturbations is a defining characteristic of classical chaos \cite{Strogatz}. 
On the other hand, it is usually assumed that there is no state sensitivity in the quantum realm \cite{Haake,Rudnicki,Peres}. 
To deal with this problem different methods to detect chaotic behavior in the quantum domain were proposed. Instead of state sensitivity, one can observe sensitivity to Hamiltonian perturbation \cite{Peres} which can be measured e.g. by Loschmidt echo \cite{Pastawski,GORIN200633,PG_2021}. A related concept to Loschmidt echo is known as the out-of-time-order correlator (OTOC) \cite{Losch,Swingle,PRLGarcia,Anand}. Also, commonly used methods to detect quantum chaos consist of examining statistics of energy levels and eigenstates \cite{Brody,Dalesio,Borgo,Wang}. Finally, it is also worth to mention entanglement dynamics as yet another powerful tool to investigate this problem \cite{Wangx,Lewy,Lerose,Gietka}.

In this Letter, contrary to common belief, we show that it is still possible to define quantum chaos with the primordial concept of quickly growing distance between the original and the perturbed state. This possibility is based on two observations. First, quantum states are not represented as points in phase space, but as vectors in Hilbert space. Second, a scalar product (overlap) between quantum states is invariant under unitary dynamics and consequently, no sensitivity to perturbation can be detected through any metric based on it. However, one can find a different metric between quantum states, which is not invariant under unitary dynamics. Here we use a slight modification of the metric proposed by Girolami and Anza \cite{Girolami} which can be called quantum Hamming distance (QHD). We apply QHD to the multi-qubit implementation of the quantum kicked top. We choose it because QHD is extremely well adjusted to systems living in Hilbert space possessing natural tensor product structure. In addition, the kicked top is a very well-known system \cite{Haake,PG_2021,Lerose,Kus,Mkus,Gerwin,Lombardi,Wangx,Ghose,Madhok,Bhosale,Chaudhury,Lewy,Neill,Dogra,Zarum,HaakeKus,GhoseS,Kumari,Fiderer,Sieberer} and our goal is not so much to study its behavior as to prove that divergence of states in Hilbert space, as measured by QHD, can provide a reliable description of chaotic dynamics.

{\it Quantum Hamming Distances.} Let us emphasize that the need for no 
overlap-based metric is not limited to the field of quantum chaos. An important concern regarding the overlap metric is its inability to detect differences between states that are physically significant. Let's consider as an example three states of n qubits
\begin{eqnarray}
|\psi_1\rangle &=& |000\ldots 00\rangle, \label{psi1} \\
|\psi_2\rangle &=& |000\ldots 01\rangle, \label{psi2} \\
|\psi_3\rangle &=& |111\ldots 11\rangle. \label{psi3}
\end{eqnarray}
Due to their mutual orthogonality, the overlap metric yields $dist(\psi_1,\psi_2)=dist(\psi_1,\psi_3)=dist(\psi_2,\psi_3)$. However, the states $|\psi_1\rangle$ and $|\psi_2\rangle$ bare much more physical resemblance than $|\psi_1\rangle$ and $|\psi_3\rangle$, or $|\psi_2\rangle$ and $|\psi_3\rangle$. In particular, an initial microscopic perturbation 
\begin{equation}\label{micro}
|\psi_1\rangle \rightarrow |\delta_1\rangle = \sqrt{1-\delta} |\psi_1\rangle + \sqrt{\delta} |\psi_2\rangle 
\end{equation}
can unitarily evolve into a macroscopically different state 
\begin{equation}\label{macro}
|\delta_2\rangle = \sqrt{1-\delta} |\psi_1\rangle + \sqrt{\delta} |\psi_3\rangle.  
\end{equation}
The overlap metric fails to capture the aforementioned micro-macro differences, which can be crucial in quantum chaotic systems. Put simply, the overlap metric quantifies how well two quantum states can be distinguished, and in the case of states (\ref{psi1}-\ref{psi3}), they are perfectly distinguishable. However, many important properties of these states extend far beyond mere distinguishability.

Let us consider quantum state metrics that can be interpreted as QHDs. 
The fundamental idea behind QHD (given by Girolami and Anza \cite{Girolami}) involves a partition, denoted as $P$, which divides the system into parts labeled as $a=1, 2,..., a_{max}$ ($1 \leq a_{max} \leq n $). Each part contains $k_{a}$ elements, hence $\sum^{a_{max}}_{a=1}k_{a}=n$.
Let $\rho$ and $\sigma$ represent two states of an $n$-partite system.
The QHD between  $\rho$ and $\sigma$ is defined \cite{Girolami} as
\begin{eqnarray}
D(\rho,\sigma) &=& \max_{P} \delta_{P}(\rho,\sigma), \label{WBL} \\ 
\delta_{P}(\rho,\sigma) &=& \sum_{a} \frac{1}{k_{a}} d(\rho_a,\sigma_a), \label{WBL2}
\end{eqnarray}
where $\rho_a$ and $\sigma_a$ are the states of subsystems with respect to a given partition $P$. In the above $d(.,.)$ stands for any metric. The definition of the QHD, specifically Eq. (\ref{WBL2}), relies on the property that the sum of metrics is also a metric. It can be observed that if $d(\rho,\sigma)$ and $\tilde{d}(\tilde{\rho},\tilde{\sigma})$ are two metrics, then $D(\rho\otimes\tilde{\rho},\sigma\otimes\tilde{\sigma}) = d(\rho,\sigma) + \tilde{d}(\tilde{\rho},\tilde{\sigma})$ is also a metric. All of the metric properties of $D(\rho\otimes\tilde{\rho},\sigma\otimes\tilde{\sigma})$ can be straightforwardly derived from the metric properties of $d(\rho,\sigma)$ and $\tilde{d}(\tilde{\rho},\tilde{\sigma})$. For instance, the triangle inequality corresponding to $D(\rho\otimes\tilde{\rho},\sigma\otimes\tilde{\sigma})$ is simply the sum of the two original triangle inequalities. This generalizes easily to the sum of an arbitrary number of metrics. 

Note that $\delta_{P}(\rho,\sigma)$ is not a metric. This is because for some partition $P$ two different states $\rho\neq \sigma$ may give rise to $\rho_a=\sigma_{a}$ for all $a$, which implies $\delta_{P}(\rho,\sigma)=0$. But metric should be equal to zero iff  $\rho = \sigma$. This is the reason why maximization is used in the definition (\ref{WBL}).

In the original version, Girolami and Anza \cite{Girolami} use Bures length (also known as Bures angle) \cite{Zyczkowski}
\begin{equation}
d(\rho,\sigma) = \cos^{-1} \Big( \text{Tr} \sqrt{ \sigma^{1/2} \rho \ \sigma^{1/2} } \Big).
\end{equation}
However, in the subsequent sections of this work, we employ the distance metric
based on trace distance \cite{Zyczkowski,nielsen}
\begin{equation}
d(\rho,\sigma) = \frac{1}{2}{\text Tr}|\rho-\sigma|.
\end{equation}
The choice of using the trace distance is motivated by its numerical tractability, as it is generally easier to evaluate computationally (especially when two states are almost identical) compared to the Bures length. With this choice, one obtains for our exemplary states (\ref{psi1}-\ref{psi3})
\begin{equation}
D(\psi_1,\psi_2) =1,
\end{equation}
\begin{equation}
D(\psi_1,\psi_3) =n
\end{equation}
and
\begin{equation}
D(\psi_2,\psi_3) =n-1.
\end{equation}
This is why we called this measure QHD.

Additionally, to streamline our analysis, we note that for any partition $P$, QHD is lower bounded by $\delta_P(\rho,\sigma)$
\begin{equation}
\delta_P(\rho,\sigma) \leq D(\rho,\sigma).  
\end{equation}
Therefore, if $\delta_P(\rho,\sigma)$ is hypersensitive to perturbation, then so is $D(\rho,\sigma)$.  

Recently, one of us demonstrated in \cite{Kurzynski} that the QHD can detect state-sensitivity in specific quantum dynamics involving three-body interactions. However, to establish the effectiveness of such metrics in capturing quantum chaotic features, it is crucial to demonstrate their ability to detect state-sensitivity in more realistic systems, particularly those already known to exhibit quantum chaos based on other criteria.

In this study, we investigate the dynamics of a quantum kicked top system consisting of $n$ qubits \cite{Wangx,Neill}, with an interaction energy parameterized as $\alpha$. We examine how a small initial perturbation evolves over time in this system. It is well-established that as the value of $\alpha$ increases, the system undergoes a quantum order-to-chaos transition \cite{Haake,HaakeKus}. To quantify the effect of perturbations, we use QHD. Our analysis reveals that QHD effectively detects hypersensitivity to perturbations for $\alpha>3$. This result establishes a meaningful connection between classical and quantum chaotic behaviors, shedding light on the intuitive links between the two.


{\it Kicked top.} The classical kicked top model describes the dynamics of an angular momentum vector $\mathbf{J}$, which is governed by the Hamiltonian $H(t) = H_0 + H_1 \sum_{n=-\infty}^{+\infty}T\delta(t-nT)$. Here, $H_0=\beta J_y$ represents the natural dynamics, and $H_1=\frac{\alpha}{2J} J_z^2$ represents the "kicks" that occur periodically with a period of $T$. It is well-established that the model undergoes a transition from order to chaos as the parameter $\alpha$ increases. This transition generally takes place in the regime $1<\alpha<6$.

Here, we examine the quantum version of the kicked top model implemented on spin $s=n/2$, which is equivalent to a system of $n$ interacting qubits. The dynamics is  discrete, and a single step of the evolution can be described by the following equation
\begin{equation}
|\psi_{t+1}\rangle = U_1 U_2 |\psi_t\rangle, 
\end{equation}
where
\begin{equation}
U_1 = e^{i\frac{\alpha}{4n} \sum_{i,j =1}^n \sigma_z^{(i)}\sigma_z^{(j)}},
\end{equation}
and
\begin{equation}
U_2 = e^{i\frac{\beta}{2} \sum_{i =1}^n \sigma_y^{(i)}}.
\end{equation}
In the above equation, $\sigma_j^{(i)}$ represents the Pauli-$j$ matrix (where $j=x,y,z$) acting on the $i$-th qubit. In physical terms, this model describes a system of spin-1/2 particles in a magnetic field with a magnitude proportional to $\beta$. The spins naturally precess, but their motion is periodically interrupted by pairwise interactions. The energy of interaction is proportional to $\alpha$. To simplify the equations of motion, it is common to choose $\beta = \pi/2$. However, even with this choice, the system can exhibit intricate behaviors. Notably, it has been demonstrated in \cite{Kus} that a quantum analog of the order-to-chaos transition can be observed within the same range of $\alpha$.


{\it Methods.} We conduct a numerical investigation into the evolution of the aforementioned system for $\beta=\pi/2$, exploring various choices of $\alpha$. In each simulation run, we initialize the system in the symmetric pure product state $\rho_0=\ket{\psi_0} \bra{\psi_0}$, where $|\psi_0\rangle = |\chi\rangle^{\otimes n}$, with $|\chi\rangle = \cos \theta |0\rangle + e^{i\phi}\sin\theta |1\rangle$. The parameters $\theta$ and $\phi$ are randomly chosen. These symmetric qubit states, known as coherent spin states \cite{Kitagawa}, are considered the most classical spin states as they minimize the uncertainty of the spin-$n/2$ operators $S_j=\frac{1}{2}\sum_{i=1}^n \sigma_j^{(i)}$. The coherent spin states are eigenstates of the spin operator $S_{\mathbf{k}}|\psi_0\rangle = \frac{n}{2}|\psi_0\rangle$, where ${\mathbf{k}}=(\cos\phi\sin\theta,\sin\phi\sin\theta,\cos\theta)$ represents the axis onto which the spin-$n/2$ is projected, $S_{\mathbf{k}}={\mathbf{k}}\cdot {\mathbf{S}}$ and ${\mathbf{S}}=(S_x,S_y,S_z)$. 

Furthermore, we examine the evolution of the perturbed (pure) state
$\rho_0\myprime=\ket{\psi_0\myprime} \bra{\psi_0\myprime}$, where $|\psi_0\myprime\rangle = |\chi^{\phantom{|}}\myprime\rangle^{\otimes n}$. $|\chi^{\phantom{|}}\myprime\rangle$ is given by $|\chi^{\phantom{|}}\myprime\rangle =R_{\varphi} |\chi\rangle$, where $R_{\varphi}=e^{i\frac{\varphi}{2}\sigma_{\mathbf{m}} }$ denotes a single-qubit rotation about a randomly chosen $\mathbf{m}$-axis, $\sigma_{\mathbf{m}}={\mathbf{m}}\cdot {\mathbf{s}}$ and ${\mathbf{s}}=(\sigma_x,\sigma_y,\sigma_z)$. The rotation angle is denoted by $\varphi$ ($\varphi \ll 1$). Finally, we evaluate the distance $D(\rho_t,\rho_t\myprime)$ and analyze its dependence on the parameters $t$, $\varphi$, and $\alpha$. 

It is important to note that both the initial state $\rho_0$ and the perturbed state $\rho_0\myprime$ are symmetric, meaning they remain unchanged under the permutation of spins. This symmetry is preserved throughout the evolution due to the symmetric nature of the evolution operators $U_1$ and $U_2$. Consequently, the states $\rho_t$ and $\rho_t\myprime$ also maintain their symmetry. As a result, the dynamics of the $n$-qubit system occur within a symmetric subspace of dimension $n+1$. This characteristic significantly simplifies the complexity of numerical simulations and facilitates the analysis of the obtained data. 

We found that in our numerical simulations, the optimal partition is a partition of the system into single qubits, i.e., $a_{max}=n$, $k_a=1$ for each $a$. Because the system is symmetric, we get
\begin{eqnarray}
 D(\rho_t,{\rho}_t\myprime)= \frac{n}{2}{\text Tr}|\tilde{\rho}_t-\tilde{\rho }_t \myprime|,
\end{eqnarray}
where $\tilde{\rho}_t$ is the state of a single-qubit subsystem of $\rho_t$ and $\tilde{\rho}_t\myprime$ is the state of a single-qubit subsystem of $\rho_t\myprime$. The general method of how to evaluate states of subsystems is given in Supplemental Material.

Finally, the value of $D(\rho_t,\rho_t\myprime)$ differs from one simulation to the other because it depends on the initial state that is chosen randomly. That is why we introduce average distance
\begin{equation}
{\mathcal{D}}_t=\langle D(\rho_t,\rho_t\myprime) \rangle_{\rho_0},
\end{equation}
in which we average over one hundred numerical runs. Therefore, ${\mathcal{D}}_t$ reflects a property of the system, not a property of a particular state. Note also, that for $\varphi \ll 1$ the initial distance
\begin{equation}
D(\rho_0,\rho_0\myprime)=\frac{n \, \varphi}{2}
\end{equation}
is independent of the initial state (see Supplemental Material).


\begin{figure}[t]
    \centering
    \begin{tabular}{cc}
    \includegraphics[width=0.24\textwidth]{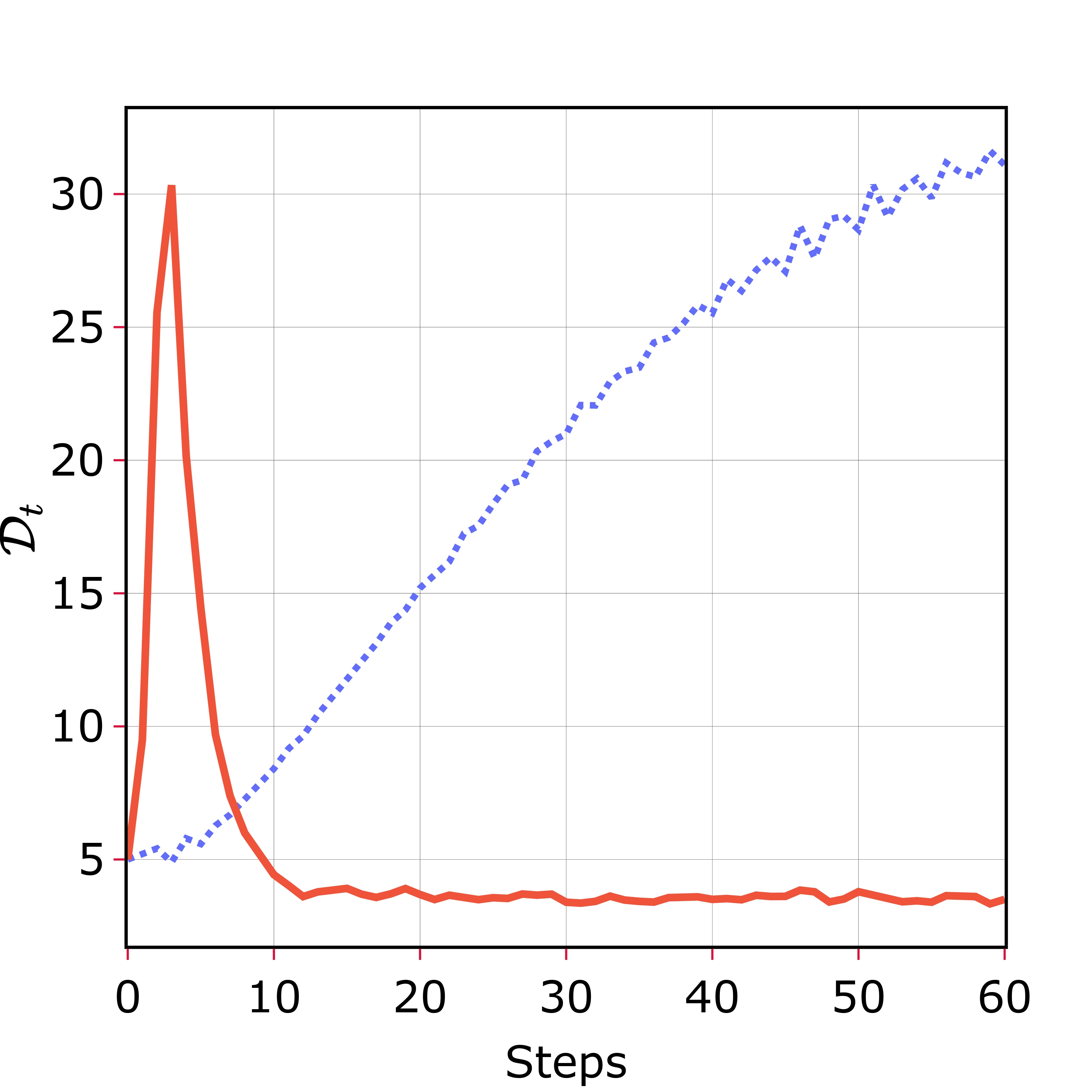}  &  \includegraphics[width=0.24\textwidth]{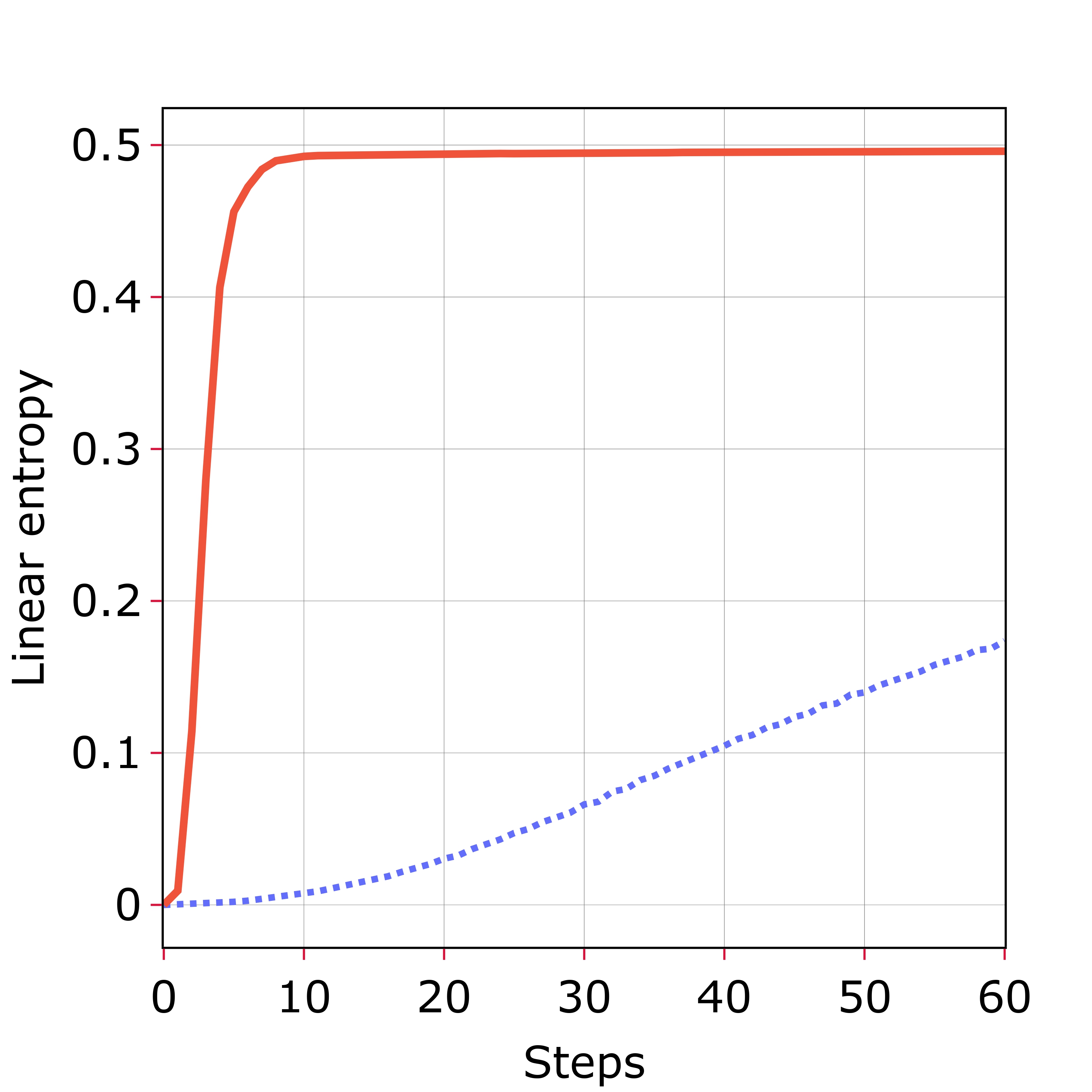}
    \end{tabular}
    \caption{Left: the plot of ${\mathcal{D}}_t$ for $n=1000$ and $\varphi=0.01$. The points were joined for better visibility. The red (solid) plot corresponds to $\alpha=6$ and the blue (dotted) one to $\alpha=1$. Right: the corresponding growth of the single-qubit entropy.}
    \label{f1}
\end{figure}

{\it Results.} Our most important finding is that ${\mathcal{D}}_t$ can be used as a witness of quantum chaos. In particular, it grows rapidly for the values of $\alpha$ corresponding to the chaotic regime and slowly for the values corresponding to the regular regime. An example of such behavior is presented in Fig. \ref{f1} (left), where we consider the system of  $n=1000$ qubits and the perturbation angle $\varphi=0.01$. We compare two cases, $\alpha=6$ (red plot) and $\alpha=1$ (blue plot). 

Interestingly, in the chaotic regime ${\mathcal{D}}_t$ grows fast and then it decreases fast, resulting in a peak. The reason for this behaviour comes from the fact that ${\mathcal{D}}_t$ is an averaged comparison between single-qubit sub-states $\tilde{\rho}_t$ and $\tilde{\rho}_t\myprime$. They initially diverge (${\mathcal{D}}_t$ grows), but due to the entangling nature of the kicked top dynamics they get more entangled with the rest of the system, therefore they become more mixed. The more mixed they become, the more similar they get (${\mathcal{D}}_t$ decreases). 

The amount of entanglement between the qubit and the rest of the system can be measured by linear entropy $S_t=1-{\text{Tr}}(\tilde{\rho}_t^{2})$. An example of how $S_t$ changes in time is represented in Fig. \ref{f1} (right). It is clear that the peak in ${\mathcal{D}}_t$ matches the growth of $S_t$. Note that the fast growth of entanglement within the multipartite system is usually considered to be a signature of quantum chaos \cite{Lerose,Wangx,Lewy,Ghose,Neill,Dogra,GhoseS,Ruebeck2017,Weinstein2006} (for detailed discussion see \cite{Kumari}). The entanglement time is analogous to the Ehrenfest time $t_E$, the time after which the correspondence principle is no longer valid. Intuitively, this happens when the uncertainties are of the order of the size of the system, which is exactly what is manifested by entangled states. 

\begin{figure}[t]
    \centering
    \begin{tabular}{cc}
    \includegraphics[width=0.24\textwidth]{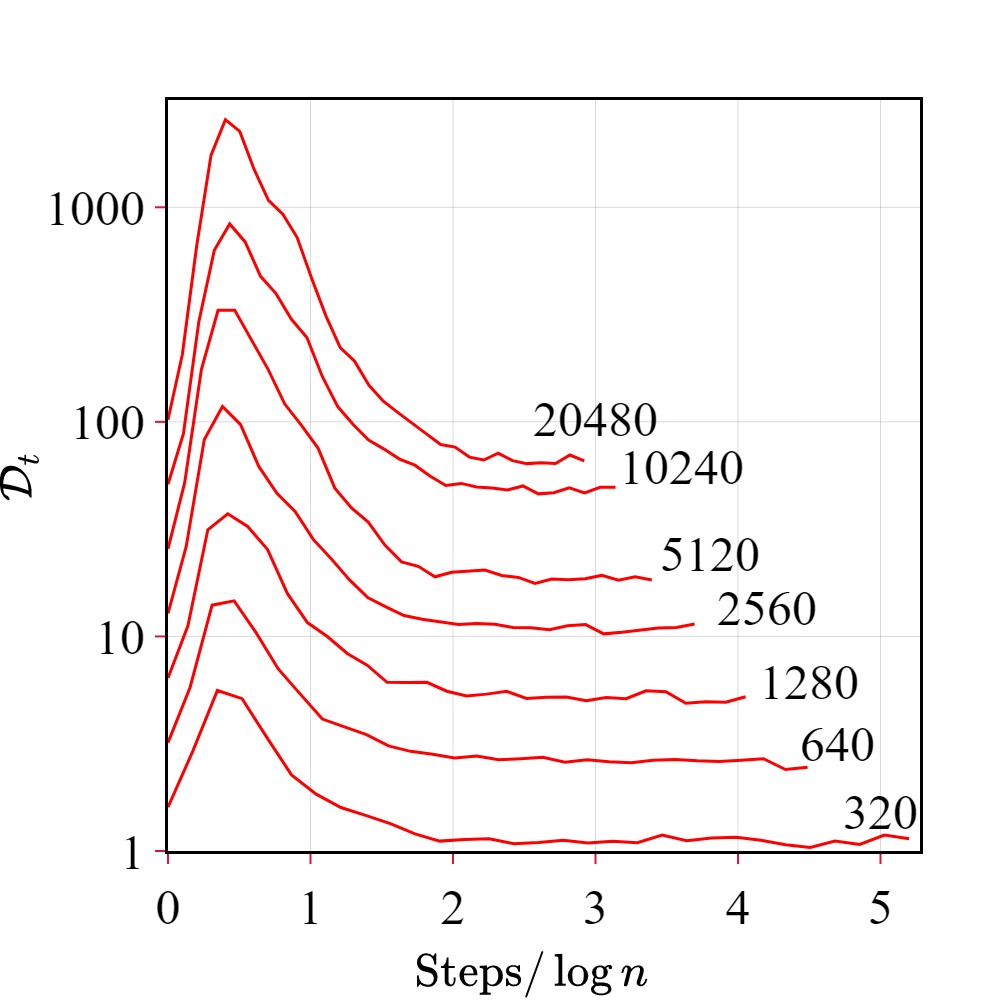}  &  \includegraphics[width=0.24\textwidth]{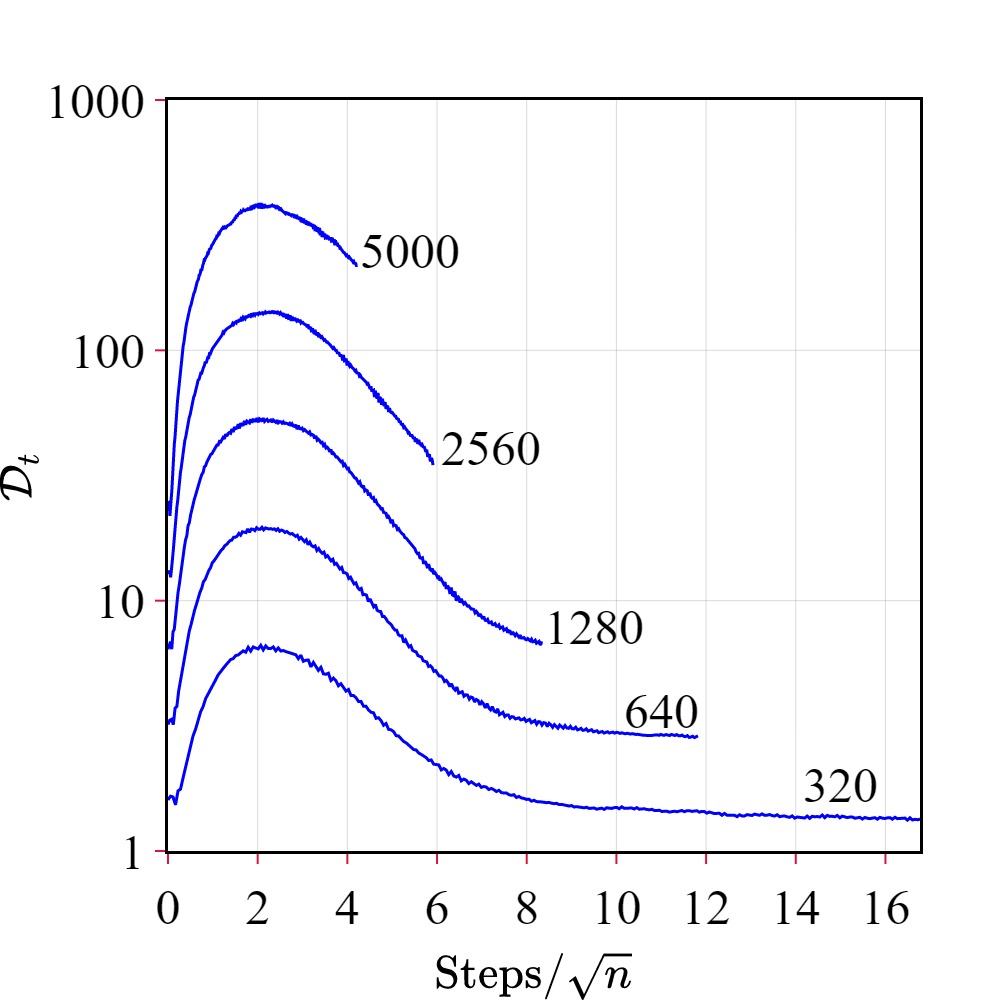}
    \end{tabular}
    \caption{The plots of ${\mathcal{D}}_t$ for $\varphi=0.01$ and different values of $n$. It is visible that the Ehrenfest time scales as $\log n$ for $\alpha=6$ and as $\sqrt{n}$ for $\alpha=1$.}
    \label{f2}
\end{figure}

Because of the above, the position of the peak of ${\mathcal{D}}_t$ can be considered as the Ehrenfest time of the system. In Fig. \ref{f2} we show how ${\mathcal{D}}_t$ depends on the number of qubits $n$. 
It is expected \cite{Lerose,Chuan,Berman} that Ehrenfest time scales as $t_E \sim h_{eff}^{-1/2}$ for regular and as $t_E \sim ln(h_{eff}^{-1})$ for chaotic dynamics, where $h_{eff}$ is the effective Planck constant. For systems confined to the symmetric subspace the effective Planck constant is \cite{Lerose,Sciolla} inversely proportional to the dimension of this subspace. Thus we expect $t_E \sim n^{1/2}$ for regular and as $t_E \sim ln(n)$ for chaotic dynamics. Position of the peaks in Fig. \ref{f2}
roughly fulfills this expectation. This confirms that our quantum chaos witness based on ${\mathcal{D}}_t$ is in accordance with the usually used ones. 

\begin{figure}[t]
    \centering
    \begin{tabular}{cc}
    \includegraphics[width=0.24\textwidth]{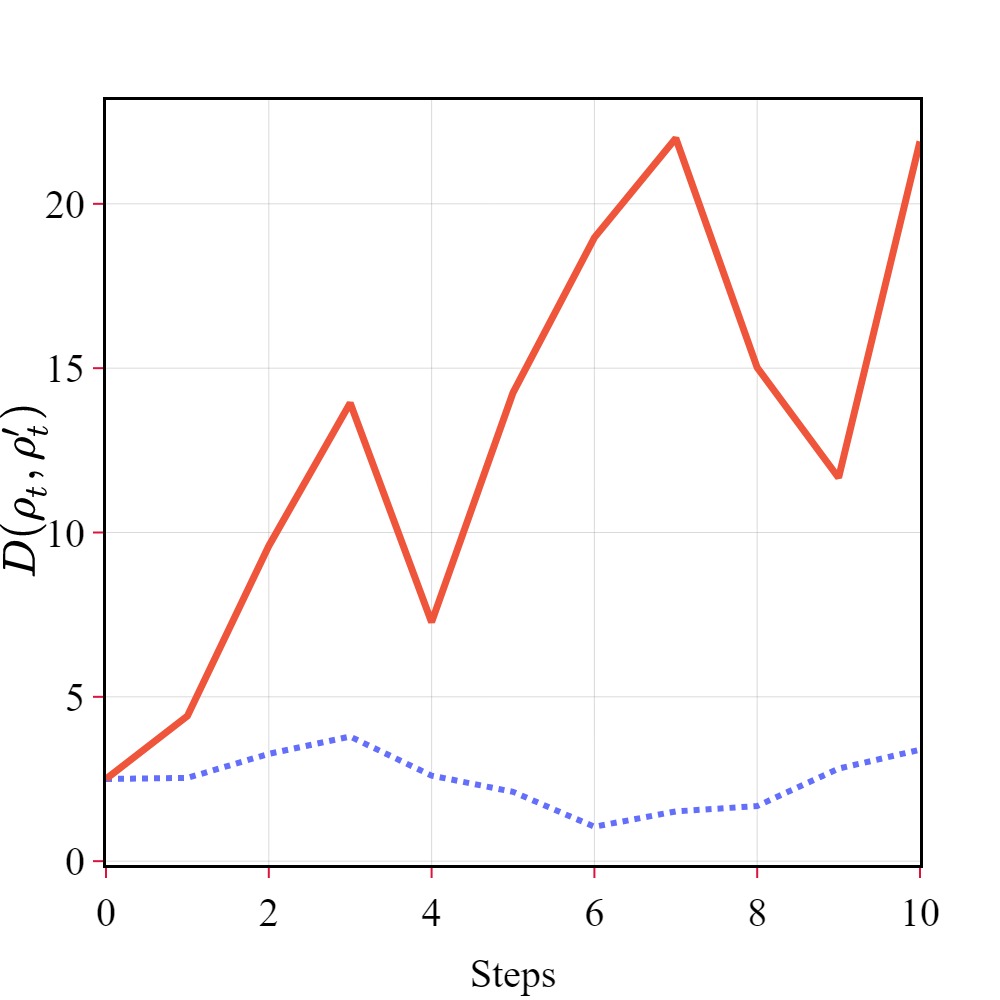}  &  \includegraphics[width=0.24\textwidth]{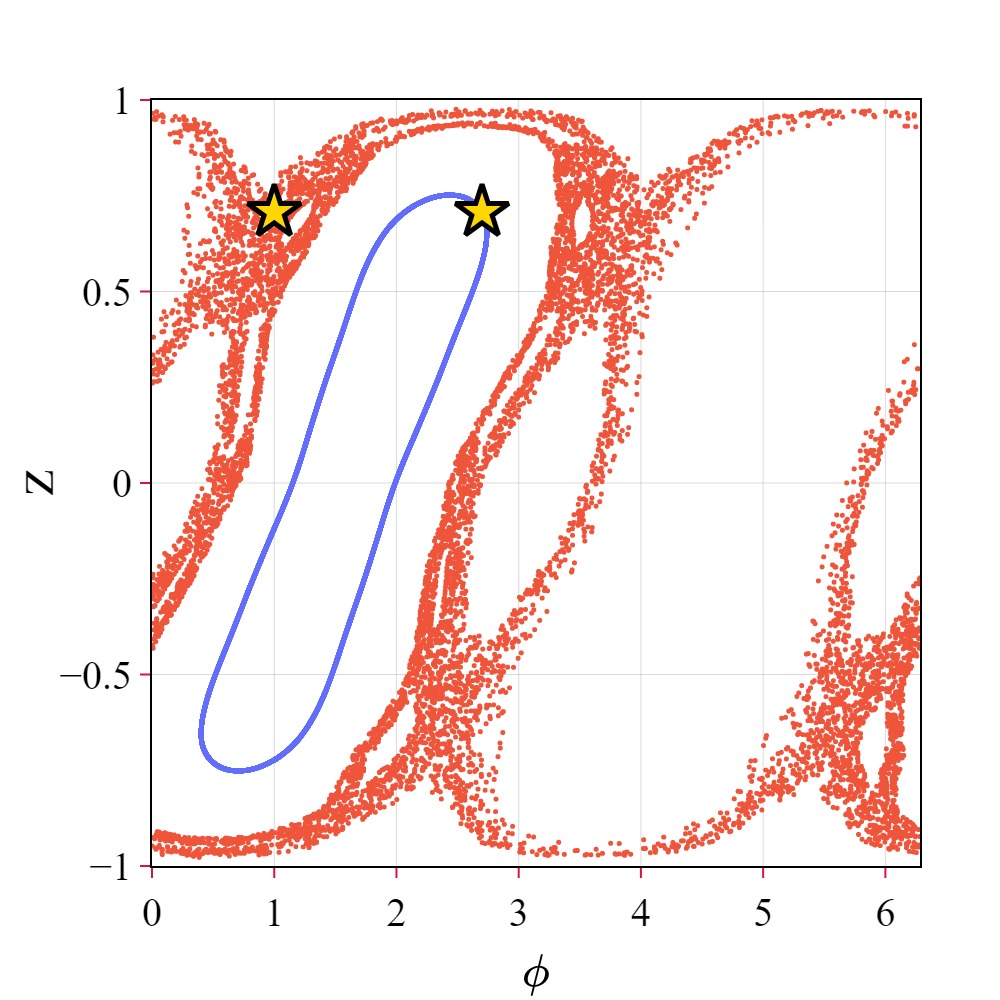}
    \end{tabular}
    \caption{Left: the plot of $D(\rho_t,\rho_t\myprime)$ for $n=500$, $\varphi=0.01$ and $\alpha=2.3$. This value of $\alpha$ corresponds to both chaotic and regular behaviors. The evolution depends on the choice of initial conditions determined by $\theta$ and $\phi$ (red (solid) and blue (dotted)). Right: The corresponding classical trajectories for the analogous initial conditions. Stars mark starting points.}
    \label{f3}
\end{figure}

Another interesting problem is to investigate if our quantum chaos witness works in the range for which both chaotic and regular behaviors are known to coexist ($1<\alpha<6$). In this case, the chaotic properties of the system depend on the initial state and we cannot use the averaged witness ${\mathcal{D}}_t$ anymore. Instead, we return to $D(\rho_t,\rho_t\myprime)$. Moreover, we relate the observed quantum behavior to the behavior of classical trajectories with analogous initial conditions. The method for evaluationg classical trajectories is given in Supplemental Material.

In Fig. \ref{f3} we show two examples for $\alpha=2.3$. In the first one (blue, dotted) the initial conditions $\theta$ and $\phi$ correspond to a classical regular trajectory (see Fig. \ref{f3} right). It is clear that in this case $D(\rho_t,\rho_t\myprime)$ does not change much (see Fig. \ref{f3} left). The second example (red, solid) corresponds to a classical chaotic trajectory and it is well visible that in this case $D(\rho_t,\rho_t\myprime)$ grows. This observation confirms that our approach could be also used to witness quantum chaos in the transition region. 


{\it Conclusions.} We showed that, contrary to common belief, quantum and classical chaos can be treated on equal footing, provided one chooses a proper metric in the system's state space. In case of classical systems one can detect chaos using Euclidean metric in phase space. Here we demonstrated that in case of quantum systems one can detect chaos using a quantum Hamming distance \cite{Girolami} (QHD) in Hilbert space. We analysed quantum kicked top dynamics of $n$ qubits and found that QHD between two quantum states can rapidly grow. This finding confirms that, just like classical chaotic dynamics, quantum chaotic dynamics is hypersensitive to perturbations. 

Our method of detecting quantum chaos is in accordance with previously developed methods. For example, we found that the growth of QHD matches the growth of entanglement \cite{Kumari}. Moreover, we argued that the time after which QHD reaches its maximum corresponds to the Ehrenfest time $t_E$ \cite{Lerose,Chuan,Berman}, which scales as $n^{1/2}$ and $\log(n)$ in the regular and chaotic regimes, respectively. Finally, we showed that QHD can be used to distinguish between regular and chaotic dynamics in situations the system exhibits both behaviours for different initial conditions.

In this work we initiate a research program of studying quantum chaos with the help of QHD. The next step is to investigate other discrete-time systems, continuous-time systems and continuous-variable systems.

{\it Acknowledgements.} We are grateful to D. Girolami, F. Anzà and M. Lewenstein for helpful discussions. This research is supported by the Polish National Science Centre (NCN) under the Maestro Grant no. DEC-2019/34/A/ST2/00081.  J.W. acknowledges support from IDUB BestStudentGRANT (NO. 010/39/UAM/0010). Part of numerical studies in this work have been carried out using resources provided by Wroclaw Centre for Networking and Supercomputing (wcss.pl), Grant No. 551 (A.S.S.).



\begin{thebibliography}{45}
\expandafter\ifx\csname natexlab\endcsname\relax\def\natexlab#1{#1}\fi
\expandafter\ifx\csname bibnamefont\endcsname\relax
  \def\bibnamefont#1{#1}\fi
\expandafter\ifx\csname bibfnamefont\endcsname\relax
  \def\bibfnamefont#1{#1}\fi
\expandafter\ifx\csname citenamefont\endcsname\relax
  \def\citenamefont#1{#1}\fi
\expandafter\ifx\csname url\endcsname\relax
  \def\url#1{\texttt{#1}}\fi
\expandafter\ifx\csname urlprefix\endcsname\relax\def\urlprefix{URL }\fi
\providecommand{\bibinfo}[2]{#2}
\providecommand{\eprint}[2][]{\url{#2}}

\bibitem[{\citenamefont{Strogatz}(2000)}]{Strogatz}
\bibinfo{author}{\bibfnamefont{S.~H.} \bibnamefont{Strogatz}},
  \bibinfo{title}{Nonlinear Dynamics And Chaos: With Applications To
  Physics, Biology, Chemistry, And Engineering (Studies in Nonlinearity)}
  (\bibinfo{publisher}{CRC Press}, \bibinfo{year}{2000}).

\bibitem[{\citenamefont{Haake et~al.}(2018)\citenamefont{Haake, Gnutzmann, and
  Kuś}}]{Haake}
\bibinfo{author}{\bibfnamefont{F.}~\bibnamefont{Haake}},
  \bibinfo{author}{\bibfnamefont{S.}~\bibnamefont{Gnutzmann}},
  \bibnamefont{and} \bibinfo{author}{\bibfnamefont{M.}~\bibnamefont{Kuś}},
  \bibinfo{journal}{{Springer Series in Synergetics}}  (\bibinfo{year}{2018}).

\bibitem[{\citenamefont{Rudnick}(2008)}]{Rudnicki}
\bibinfo{author}{\bibfnamefont{Z.}~\bibnamefont{Rudnick}},
  \bibinfo{journal}{{Notices of the American Mathematical Society}}
  \textbf{\bibinfo{volume}{55}}, \bibinfo{pages}{32} (\bibinfo{year}{2008}).

\bibitem[{\citenamefont{Peres}(1984)}]{Peres}
\bibinfo{author}{\bibfnamefont{A.}~\bibnamefont{Peres}},
  \bibinfo{journal}{Physical Review A} \textbf{\bibinfo{volume}{30}},
  \bibinfo{pages}{1610} (\bibinfo{year}{1984}).

\bibitem[{\citenamefont{Pastawski et~al.}(1995)\citenamefont{Pastawski,
  Levstein, and Usaj}}]{Pastawski}
\bibinfo{author}{\bibfnamefont{H.}~\bibnamefont{Pastawski}},
  \bibinfo{author}{\bibfnamefont{P.}~\bibnamefont{Levstein}}, \bibnamefont{and}
  \bibinfo{author}{\bibfnamefont{G.}~\bibnamefont{Usaj}},
  \bibinfo{journal}{Physical Review Letters} \textbf{\bibinfo{volume}{75}},
  \bibinfo{pages}{4310} (\bibinfo{year}{1995}).

\bibitem[{\citenamefont{Gorin et~al.}(2006)\citenamefont{Gorin, Prosen,
  Seligman, and Žnidarič}}]{GORIN200633}
\bibinfo{author}{\bibfnamefont{T.}~\bibnamefont{Gorin}},
  \bibinfo{author}{\bibfnamefont{T.}~\bibnamefont{Prosen}},
  \bibinfo{author}{\bibfnamefont{T.}~\bibnamefont{Seligman}}, \bibnamefont{and}
  \bibinfo{author}{\bibfnamefont{M.}~\bibnamefont{Žnidarič}},
  \bibinfo{journal}{Physics Reports} \textbf{\bibinfo{volume}{435}},
  \bibinfo{pages}{33} (\bibinfo{year}{2006}).

\bibitem[{\citenamefont{Pg et~al.}(2021)\citenamefont{Pg, Madhok, and
  Lakshminarayan}}]{PG_2021}
\bibinfo{author}{\bibfnamefont{S.}~\bibnamefont{Pg}},
  \bibinfo{author}{\bibfnamefont{V.}~\bibnamefont{Madhok}}, \bibnamefont{and}
  \bibinfo{author}{\bibfnamefont{A.}~\bibnamefont{Lakshminarayan}},
  \bibinfo{journal}{Journal of Physics D: Applied Physics}
  \textbf{\bibinfo{volume}{54}}, \bibinfo{pages}{274004}
  (\bibinfo{year}{2021}).

\bibitem[{\citenamefont{Yan et~al.}(2020)\citenamefont{Yan, Cincio, and
  Zurek}}]{Losch}
\bibinfo{author}{\bibfnamefont{B.}~\bibnamefont{Yan}},
  \bibinfo{author}{\bibfnamefont{L.}~\bibnamefont{Cincio}}, \bibnamefont{and}
  \bibinfo{author}{\bibfnamefont{W.}~\bibnamefont{Zurek}},
  \bibinfo{journal}{Physical Review Letters} \textbf{\bibinfo{volume}{124}}
  (\bibinfo{year}{2020}).

\bibitem[{\citenamefont{Swingle}(2018)}]{Swingle}
\bibinfo{author}{\bibfnamefont{B.}~\bibnamefont{Swingle}},
  \bibinfo{journal}{Nature Physics} \textbf{\bibinfo{volume}{14}},
  \bibinfo{pages}{988} (\bibinfo{year}{2018}).

\bibitem[{\citenamefont{García-Mata et~al.}(2018)\citenamefont{García-Mata,
  Saraceno, Jalabert, Roncaglia, and Wisniacki}}]{PRLGarcia}
\bibinfo{author}{\bibfnamefont{I.}~\bibnamefont{García-Mata}},
  \bibinfo{author}{\bibfnamefont{M.}~\bibnamefont{Saraceno}},
  \bibinfo{author}{\bibfnamefont{R.}~\bibnamefont{Jalabert}},
  \bibinfo{author}{\bibfnamefont{A.}~\bibnamefont{Roncaglia}},
  \bibnamefont{and}
  \bibinfo{author}{\bibfnamefont{D.}~\bibnamefont{Wisniacki}},
  \bibinfo{journal}{Physical Review Letters} \textbf{\bibinfo{volume}{121}},
  \bibinfo{pages}{210601} (\bibinfo{year}{2018}).

\bibitem[{\citenamefont{Anand et~al.}(2021)\citenamefont{Anand, Styliaris,
  Kumari, and Zanardi}}]{Anand}
\bibinfo{author}{\bibfnamefont{N.}~\bibnamefont{Anand}},
  \bibinfo{author}{\bibfnamefont{G.}~\bibnamefont{Styliaris}},
  \bibinfo{author}{\bibfnamefont{M.}~\bibnamefont{Kumari}}, \bibnamefont{and}
  \bibinfo{author}{\bibfnamefont{P.}~\bibnamefont{Zanardi}},
  \bibinfo{journal}{Physical Review Research} \textbf{\bibinfo{volume}{3}},
  \bibinfo{pages}{23214} (\bibinfo{year}{2021}).

\bibitem[{\citenamefont{Brody et~al.}(1981)\citenamefont{Brody, Flores, French,
  Mello, Pandey, and Wong}}]{Brody}
\bibinfo{author}{\bibfnamefont{T.~A.} \bibnamefont{Brody}},
  \bibinfo{author}{\bibfnamefont{J.}~\bibnamefont{Flores}},
  \bibinfo{author}{\bibfnamefont{J.~B.} \bibnamefont{French}},
  \bibinfo{author}{\bibfnamefont{P.~A.} \bibnamefont{Mello}},
  \bibinfo{author}{\bibfnamefont{A.}~\bibnamefont{Pandey}}, \bibnamefont{and}
  \bibinfo{author}{\bibfnamefont{S.~S.~M.} \bibnamefont{Wong}},
  \bibinfo{journal}{Rev. Mod. Phys.} \textbf{\bibinfo{volume}{53}},
  \bibinfo{pages}{385} (\bibinfo{year}{1981}),

\bibitem[{\citenamefont{D'alessio et~al.}(2016)\citenamefont{D'alessio, Kafri,
  Polkovnikov, and Rigol}}]{Dalesio}
\bibinfo{author}{\bibfnamefont{L.}~\bibnamefont{D'alessio}},
  \bibinfo{author}{\bibfnamefont{Y.}~\bibnamefont{Kafri}},
  \bibinfo{author}{\bibfnamefont{A.}~\bibnamefont{Polkovnikov}},
  \bibnamefont{and} \bibinfo{author}{\bibfnamefont{M.}~\bibnamefont{Rigol}},
  \bibinfo{journal}{Advances in Physics} \textbf{\bibinfo{volume}{65}},
  \bibinfo{pages}{239} (\bibinfo{year}{2016}).

\bibitem[{\citenamefont{Borgonovi et~al.}(2016)\citenamefont{Borgonovi,
  Izrailev, Santos, and Zelevinsky}}]{Borgo}
\bibinfo{author}{\bibfnamefont{F.}~\bibnamefont{Borgonovi}},
  \bibinfo{author}{\bibfnamefont{F.}~\bibnamefont{Izrailev}},
  \bibinfo{author}{\bibfnamefont{L.}~\bibnamefont{Santos}}, \bibnamefont{and}
  \bibinfo{author}{\bibfnamefont{V.}~\bibnamefont{Zelevinsky}},
  \bibinfo{journal}{Physics Reports} \textbf{\bibinfo{volume}{626}},
  \bibinfo{pages}{1} (\bibinfo{year}{2016}).

\bibitem[{\citenamefont{Wang et~al.}(2023)\citenamefont{Wang, Ma, Zhang, and
  Wang}}]{Wang}
\bibinfo{author}{\bibfnamefont{X.-Q.} \bibnamefont{Wang}},
  \bibinfo{author}{\bibfnamefont{J.}~\bibnamefont{Ma}},
  \bibinfo{author}{\bibfnamefont{X.-H.} \bibnamefont{Zhang}}, \bibnamefont{and}
  \bibinfo{author}{\bibfnamefont{X.-G.} \bibnamefont{Wang}},
  \bibinfo{journal}{Chinese Physics B} \textbf{\bibinfo{volume}{20}},
  \bibinfo{pages}{050510} (\bibinfo{year}{2023}).

\bibitem[{\citenamefont{Wang et~al.}(2004)\citenamefont{Wang, Ghose, Sanders,
  and Hu}}]{Wangx}
\bibinfo{author}{\bibfnamefont{X.}~\bibnamefont{Wang}},
  \bibinfo{author}{\bibfnamefont{S.}~\bibnamefont{Ghose}},
  \bibinfo{author}{\bibfnamefont{B.}~\bibnamefont{Sanders}}, \bibnamefont{and}
  \bibinfo{author}{\bibfnamefont{B.}~\bibnamefont{Hu}},
  \bibinfo{journal}{Physical Review E} \textbf{\bibinfo{volume}{70}},
  \bibinfo{pages}{16217} (\bibinfo{year}{2004}).

\bibitem[{\citenamefont{Piga et~al.}(2019)\citenamefont{Piga, Lewenstein, and
  Quach}}]{Lewy}
\bibinfo{author}{\bibfnamefont{A.}~\bibnamefont{Piga}},
  \bibinfo{author}{\bibfnamefont{M.}~\bibnamefont{Lewenstein}},
  \bibnamefont{and} \bibinfo{author}{\bibfnamefont{J.}~\bibnamefont{Quach}},
  \bibinfo{journal}{Physical Review E} \textbf{\bibinfo{volume}{99}},
  \bibinfo{pages}{32213} (\bibinfo{year}{2019}).

\bibitem[{\citenamefont{Lerose and Pappalardi}(2020)}]{Lerose}
\bibinfo{author}{\bibfnamefont{A.}~\bibnamefont{Lerose}} \bibnamefont{and}
  \bibinfo{author}{\bibfnamefont{S.}~\bibnamefont{Pappalardi}},
  \bibinfo{journal}{Physical Review A} \textbf{\bibinfo{volume}{102}},
  \bibinfo{pages}{32404} (\bibinfo{year}{2020}).

\bibitem[{\citenamefont{Gietka et~al.}(2019)\citenamefont{Gietka, Chwedeńczuk,
  Wasak, and Piazza}}]{Gietka}
\bibinfo{author}{\bibfnamefont{K.}~\bibnamefont{Gietka}},
  \bibinfo{author}{\bibfnamefont{J.}~\bibnamefont{Chwedeńczuk}},
  \bibinfo{author}{\bibfnamefont{T.}~\bibnamefont{Wasak}}, \bibnamefont{and}
  \bibinfo{author}{\bibfnamefont{F.}~\bibnamefont{Piazza}},
  \bibinfo{journal}{Physical Review B} \textbf{\bibinfo{volume}{99}},
  \bibinfo{pages}{64303} (\bibinfo{year}{2019}).

\bibitem[{\citenamefont{Girolami and Anzà}(2021)}]{Girolami}
\bibinfo{author}{\bibfnamefont{D.}~\bibnamefont{Girolami}} \bibnamefont{and}
  \bibinfo{author}{\bibfnamefont{F.}~\bibnamefont{Anzà}},
  \bibinfo{journal}{Physical Review Letters} \textbf{\bibinfo{volume}{126}},
  \bibinfo{pages}{170502} (\bibinfo{year}{2021}).

\bibitem[{\citenamefont{Kuś et~al.}(1987)\citenamefont{Kuś, Scharf, and
  Haake}}]{Kus}
\bibinfo{author}{\bibfnamefont{M.}~\bibnamefont{Kuś}},
  \bibinfo{author}{\bibfnamefont{R.}~\bibnamefont{Scharf}}, \bibnamefont{and}
  \bibinfo{author}{\bibfnamefont{F.}~\bibnamefont{Haake}},
  \bibinfo{journal}{Zeitschrift für Physik B Condensed Matter}
  \textbf{\bibinfo{volume}{66}}, \bibinfo{pages}{129} (\bibinfo{year}{1987}),
  ISSN \bibinfo{issn}{1431-584X},

\bibitem[{\citenamefont{Kus et~al.}(1988)\citenamefont{Kus, Mostowski, and
  Haake}}]{Mkus}
\bibinfo{author}{\bibfnamefont{M.}~\bibnamefont{Kus}},
  \bibinfo{author}{\bibfnamefont{J.}~\bibnamefont{Mostowski}},
  \bibnamefont{and} \bibinfo{author}{\bibfnamefont{F.}~\bibnamefont{Haake}},
  \bibinfo{journal}{Journal of Physics A: Mathematical and General}
  \textbf{\bibinfo{volume}{21}}, \bibinfo{pages}{L1073} (\bibinfo{year}{1988}).

\bibitem[{\citenamefont{Gerwinski et~al.}(1995)\citenamefont{Gerwinski, Haake,
  Wiedemann, Kuś, and Życzkowski}}]{Gerwin}
\bibinfo{author}{\bibfnamefont{P.}~\bibnamefont{Gerwinski}},
  \bibinfo{author}{\bibfnamefont{F.}~\bibnamefont{Haake}},
  \bibinfo{author}{\bibfnamefont{H.}~\bibnamefont{Wiedemann}},
  \bibinfo{author}{\bibfnamefont{M.}~\bibnamefont{Kuś}}, \bibnamefont{and}
  \bibinfo{author}{\bibfnamefont{K.}~\bibnamefont{Życzkowski}},
  \bibinfo{journal}{Physical Review Letters} \textbf{\bibinfo{volume}{74}},
  \bibinfo{pages}{1562} (\bibinfo{year}{1995}).

\bibitem[{\citenamefont{Lombardi and Matzkin}(2011)}]{Lombardi}
\bibinfo{author}{\bibfnamefont{M.}~\bibnamefont{Lombardi}} \bibnamefont{and}
  \bibinfo{author}{\bibfnamefont{A.}~\bibnamefont{Matzkin}},
  \bibinfo{journal}{Physical Review E} \textbf{\bibinfo{volume}{83}},
  \bibinfo{pages}{16207} (\bibinfo{year}{2011}).

\bibitem[{\citenamefont{Ghose and Sanders}(2004)}]{Ghose}
\bibinfo{author}{\bibfnamefont{S.}~\bibnamefont{Ghose}} \bibnamefont{and}
  \bibinfo{author}{\bibfnamefont{B.}~\bibnamefont{Sanders}},
  \bibinfo{journal}{Physical Review A} \textbf{\bibinfo{volume}{70}},
  \bibinfo{pages}{62315} (\bibinfo{year}{2004}).

\bibitem[{\citenamefont{Madhok et~al.}(2014)\citenamefont{Madhok, Riofrío,
  Ghose, and Deutsch}}]{Madhok}
\bibinfo{author}{\bibfnamefont{V.}~\bibnamefont{Madhok}},
  \bibinfo{author}{\bibfnamefont{C.}~\bibnamefont{Riofrío}},
  \bibinfo{author}{\bibfnamefont{S.}~\bibnamefont{Ghose}}, \bibnamefont{and}
  \bibinfo{author}{\bibfnamefont{I.}~\bibnamefont{Deutsch}},
  \bibinfo{journal}{Physical Review Letters} \textbf{\bibinfo{volume}{112}},
  \bibinfo{pages}{14102} (\bibinfo{year}{2014}).

\bibitem[{\citenamefont{Bhosale and Santhanam}(2017)}]{Bhosale}
\bibinfo{author}{\bibfnamefont{U.~T.} \bibnamefont{Bhosale}} \bibnamefont{and}
  \bibinfo{author}{\bibfnamefont{M.~S.} \bibnamefont{Santhanam}},
  \bibinfo{journal}{Phys. Rev. E} \textbf{\bibinfo{volume}{95}},
  \bibinfo{pages}{012216} (\bibinfo{year}{2017}),

\bibitem[{\citenamefont{Chaudhury et~al.}(2009)\citenamefont{Chaudhury, Smith,
  Anderson, Ghose, and Jessen}}]{Chaudhury}
\bibinfo{author}{\bibfnamefont{S.}~\bibnamefont{Chaudhury}},
  \bibinfo{author}{\bibfnamefont{A.}~\bibnamefont{Smith}},
  \bibinfo{author}{\bibfnamefont{B.}~\bibnamefont{Anderson}},
  \bibinfo{author}{\bibfnamefont{S.}~\bibnamefont{Ghose}}, \bibnamefont{and}
  \bibinfo{author}{\bibfnamefont{P.}~\bibnamefont{Jessen}},
  \bibinfo{journal}{Nature} \textbf{\bibinfo{volume}{461}},
  \bibinfo{pages}{768} (\bibinfo{year}{2009}).

\bibitem[{\citenamefont{Neill et~al.}(2016)\citenamefont{Neill, Roushan, Fang,
  Chen, Kolodrubetz, Chen, Megrant, Barends, Campbell, Chiaro et~al.}}]{Neill}
\bibinfo{author}{\bibfnamefont{C.}~\bibnamefont{Neill}},
  \bibinfo{author}{\bibfnamefont{P.}~\bibnamefont{Roushan}},
  \bibinfo{author}{\bibfnamefont{M.}~\bibnamefont{Fang}},
  \bibinfo{author}{\bibfnamefont{Y.}~\bibnamefont{Chen}},
  \bibinfo{author}{\bibfnamefont{M.}~\bibnamefont{Kolodrubetz}},
  \bibinfo{author}{\bibfnamefont{Z.}~\bibnamefont{Chen}},
  \bibinfo{author}{\bibfnamefont{A.}~\bibnamefont{Megrant}},
  \bibinfo{author}{\bibfnamefont{R.}~\bibnamefont{Barends}},
  \bibinfo{author}{\bibfnamefont{B.}~\bibnamefont{Campbell}},
  \bibinfo{author}{\bibfnamefont{B.}~\bibnamefont{Chiaro}},
  \bibnamefont{et~al.}, \bibinfo{journal}{Nature Physics}
  \textbf{\bibinfo{volume}{12}}, \bibinfo{pages}{1037} (\bibinfo{year}{2016}).

\bibitem[{\citenamefont{Dogra et~al.}(2019)\citenamefont{Dogra, Madhok, and
  Lakshminarayan}}]{Dogra}
\bibinfo{author}{\bibfnamefont{S.}~\bibnamefont{Dogra}},
  \bibinfo{author}{\bibfnamefont{V.}~\bibnamefont{Madhok}}, \bibnamefont{and}
  \bibinfo{author}{\bibfnamefont{A.}~\bibnamefont{Lakshminarayan}},
  \bibinfo{journal}{Physical Review E} \textbf{\bibinfo{volume}{99}},
  \bibinfo{pages}{062217} (\bibinfo{year}{2019}).

\bibitem[{\citenamefont{Zarum and Sarkar}(1998)}]{Zarum}
\bibinfo{author}{\bibfnamefont{R.}~\bibnamefont{Zarum}} \bibnamefont{and}
  \bibinfo{author}{\bibfnamefont{S.}~\bibnamefont{Sarkar}},
  \bibinfo{journal}{Physical Review E} \textbf{\bibinfo{volume}{57}},
  \bibinfo{pages}{5467} (\bibinfo{year}{1998}).

\bibitem[{\citenamefont{Haake et~al.}(1987)\citenamefont{Haake, Kuś, and
  Scharf}}]{HaakeKus}
\bibinfo{author}{\bibfnamefont{F.}~\bibnamefont{Haake}},
  \bibinfo{author}{\bibfnamefont{M.}~\bibnamefont{Kuś}}, \bibnamefont{and}
  \bibinfo{author}{\bibfnamefont{R.}~\bibnamefont{Scharf}},
  \bibinfo{journal}{Zeitschrift für Physik B Condensed Matter}
  \textbf{\bibinfo{volume}{65}}, \bibinfo{pages}{381} (\bibinfo{year}{1987}).

\bibitem[{\citenamefont{Ghose et~al.}(2008)\citenamefont{Ghose, Stock, Jessen,
  Lal, and Silberfarb}}]{GhoseS}
\bibinfo{author}{\bibfnamefont{S.}~\bibnamefont{Ghose}},
  \bibinfo{author}{\bibfnamefont{R.}~\bibnamefont{Stock}},
  \bibinfo{author}{\bibfnamefont{P.}~\bibnamefont{Jessen}},
  \bibinfo{author}{\bibfnamefont{R.}~\bibnamefont{Lal}}, \bibnamefont{and}
  \bibinfo{author}{\bibfnamefont{A.}~\bibnamefont{Silberfarb}},
  \bibinfo{journal}{Physical Review A} \textbf{\bibinfo{volume}{78}},
  \bibinfo{pages}{042318} (\bibinfo{year}{2008}).

\bibitem[{\citenamefont{Kumari and Ghose}(2019)}]{Kumari}
\bibinfo{author}{\bibfnamefont{M.}~\bibnamefont{Kumari}} \bibnamefont{and}
  \bibinfo{author}{\bibfnamefont{S.}~\bibnamefont{Ghose}},
  \bibinfo{journal}{Physical Review A} \textbf{\bibinfo{volume}{99}},
  \bibinfo{pages}{042311} (\bibinfo{year}{2019}).

\bibitem[{\citenamefont{Fiderer and Braun}(2018)}]{Fiderer}
\bibinfo{author}{\bibfnamefont{L.}~\bibnamefont{Fiderer}} \bibnamefont{and}
  \bibinfo{author}{\bibfnamefont{D.}~\bibnamefont{Braun}},
  \bibinfo{journal}{Nature Communications} \textbf{\bibinfo{volume}{9}},
  \bibinfo{pages}{1351} (\bibinfo{year}{2018}).

\bibitem[{\citenamefont{Sieberer et~al.}(2019)\citenamefont{Sieberer, Olsacher,
  Elben, Heyl, Hauke, Haake, and Zoller}}]{Sieberer}
\bibinfo{author}{\bibfnamefont{L.}~\bibnamefont{Sieberer}},
  \bibinfo{author}{\bibfnamefont{T.}~\bibnamefont{Olsacher}},
  \bibinfo{author}{\bibfnamefont{A.}~\bibnamefont{Elben}},
  \bibinfo{author}{\bibfnamefont{M.}~\bibnamefont{Heyl}},
  \bibinfo{author}{\bibfnamefont{P.}~\bibnamefont{Hauke}},
  \bibinfo{author}{\bibfnamefont{F.}~\bibnamefont{Haake}}, \bibnamefont{and}
  \bibinfo{author}{\bibfnamefont{P.}~\bibnamefont{Zoller}},
  \bibinfo{journal}{npj Quantum Information} \textbf{\bibinfo{volume}{5}},
  \bibinfo{pages}{78} (\bibinfo{year}{2019}).

\bibitem[{\citenamefont{Bengtsson and Życzkowski}(2006)}]{Zyczkowski}
\bibinfo{author}{\bibfnamefont{I.}~\bibnamefont{Bengtsson}} \bibnamefont{and}
  \bibinfo{author}{\bibfnamefont{K.}~\bibnamefont{Życzkowski}},
  \bibinfo{title}{Geometry of quantum states} (\bibinfo{year}{2006}).

\bibitem[{\citenamefont{Nielsen and Chuang}(2010)}]{nielsen}
\bibinfo{author}{\bibfnamefont{M.}~\bibnamefont{Nielsen}} \bibnamefont{and}
  \bibinfo{author}{\bibfnamefont{I.}~\bibnamefont{Chuang}}
  (\bibinfo{publisher}{Cambridge University Press}, \bibinfo{year}{2010}).

\bibitem[{\citenamefont{Kurzy\ifmmode~\acute{n}\else
  \'{n}\fi{}ski}(2021)}]{Kurzynski}
\bibinfo{author}{\bibfnamefont{P.}~\bibnamefont{Kurzy\ifmmode~\acute{n}\else
  \'{n}\fi{}ski}}, \bibinfo{journal}{Phys. Rev. E}
  \textbf{\bibinfo{volume}{104}}, \bibinfo{pages}{L052202}
  (\bibinfo{year}{2021}),

\bibitem[{\citenamefont{Kitagawa and Ueda}(1993)}]{Kitagawa}
\bibinfo{author}{\bibfnamefont{M.}~\bibnamefont{Kitagawa}} \bibnamefont{and}
  \bibinfo{author}{\bibfnamefont{M.}~\bibnamefont{Ueda}},
  \bibinfo{journal}{Physical Review A} \textbf{\bibinfo{volume}{47}},
  \bibinfo{pages}{5138} (\bibinfo{year}{1993}).

\bibitem[{\citenamefont{Ruebeck et~al.}(2017)\citenamefont{Ruebeck, Lin, and
  Pattanayak}}]{Ruebeck2017}
\bibinfo{author}{\bibfnamefont{J.~B.} \bibnamefont{Ruebeck}},
  \bibinfo{author}{\bibfnamefont{J.}~\bibnamefont{Lin}}, \bibnamefont{and}
  \bibinfo{author}{\bibfnamefont{A.~K.} \bibnamefont{Pattanayak}},
  \bibinfo{journal}{Phys. Rev. E} \textbf{\bibinfo{volume}{95}},
  \bibinfo{pages}{062222} (\bibinfo{year}{2017}),

\bibitem[{\citenamefont{Weinstein and Viola}(2006)}]{Weinstein2006}
\bibinfo{author}{\bibfnamefont{Y.}~\bibnamefont{Weinstein}} \bibnamefont{and}
  \bibinfo{author}{\bibfnamefont{L.}~\bibnamefont{Viola}},
  \bibinfo{journal}{Europhysics Letters (EPL)} \textbf{\bibinfo{volume}{76}},
  \bibinfo{pages}{746} (\bibinfo{year}{2006}).

\bibitem[{\citenamefont{Chuan and Biao}(2021)}]{Chuan}
\bibinfo{author}{\bibfnamefont{Z.}~\bibnamefont{Chuan}} \bibnamefont{and}
  \bibinfo{author}{\bibnamefont{Biao}}, \bibinfo{journal}{Wu. {Chin. Phys.
  Lett.}} \textbf{\bibinfo{volume}{38}}, \bibinfo{pages}{30502}
  (\bibinfo{year}{2021}).

\bibitem[{\citenamefont{Berman et~al.}(1994)\citenamefont{Berman, Bulgakov,
  Holm, and Crossover}}]{Berman}
\bibinfo{author}{\bibfnamefont{G.}~\bibnamefont{Berman}},
  \bibinfo{author}{\bibfnamefont{E.}~\bibnamefont{Bulgakov}},
  \bibinfo{author}{\bibfnamefont{D.}~\bibnamefont{Holm}}, 
  \bibinfo{title}{\bibnamefont{Crossover-Time in Quantum Boson and Spin Systems,}} 
  \bibinfo{title}{\bibnamefont{Springer-Verlag}} 
  (\bibinfo{year}{1994}).

\bibitem[{\citenamefont{Sciolla and Biroli}(2011)}]{Sciolla}
\bibinfo{author}{\bibfnamefont{B.}~\bibnamefont{Sciolla}} \bibnamefont{and}
  \bibinfo{author}{\bibfnamefont{G.}~\bibnamefont{Biroli}},
  \bibinfo{journal}{Journal of Statistical Mechanics: Theory and Experiment}
  \textbf{\bibinfo{volume}{2011}}, \bibinfo{pages}{P11003}
  (\bibinfo{year}{2011}),

\end{thebibliography}




\section{Supplemental Material}


\subsection*{Calculation of m-qubit density matrix}

As mentioned in the main text, the evolution operators are symmetric. Thus, evolution will be constrained to the $n+1$ dimensional effective Hilbert space, provided that the system will be initialized in a symmetric state. This effective Hilbert space is spanned by the Dicke states \cite{}
\begin{equation}
    \ket{d_k^n}=\binom{n}{k}^{-1/2}\ket{n,k}+permutations.
\end{equation}
In above we use particular states  of of $n$ qubits with $k$ ones and $n-k$ zeros 
\begin{equation}
\ket{n,k}=\ket{1...10...0}=\ket{1}^{\otimes k}\otimes \ket{0}^{\otimes n-k},
\end{equation}

Now, let us assume that the $n$-qubit system is in the arbitrary symmetric state
\begin{equation}
    \ket{\Psi}=\sum_{k=0}^n a_k \ket{d_k^n},
\end{equation}
\begin{equation}
    \rho=\ket{\Psi}\bra{\Psi}=\sum_{k,k'=0}^n a_k a_{k'}^* \ket{d_k^n}\bra{d_{k'}^n}.
\end{equation}
We need to consider the subsets of $m$ qubits achieved via tracing over $n-m$ qubits. The corresponding $m$-qubit subsystem state ($m\leq n$) is given by
\begin{equation}
    \rho^{(m)}=\sum_{l=0}^n \bra{n-m,l} \rho \ket{n-m,l}+permutations,
\end{equation}
It follows that (for $l\leq k$)
\begin{eqnarray}
   \braket{n-m,l}{d_k^n} &=& \binom{n}{k}^{-1/2}\ket{1}^{\otimes (k-l)}\otimes \ket{0}^{\otimes m-(k-l)} \nonumber \\
   &+& permutations.
\end{eqnarray}
For $l> k$ one has $\braket{n-m,l}{d_k^n}=0$.
The same result is obtained for any of $\binom{n-m}{l}$ permutations of $ \ket{n-m,l}$. The above equation (for $l\leq k$) can be also written as
\begin{equation}
   \braket{n-m,l}{d_k^n} =\binom{n}{k}^{-1/2}\binom{m}{k-l}^{1/2}\ket{d_{k-l}^{m}}.
\end{equation}
Therefore
\begin{eqnarray}\nonumber
    \rho^{(m)}=\sum_{k,k'=0}^n a_k a_{k'}^* \binom{n}{k}^{-1/2} \binom{n}{k'}^{-1/2}\times \\
    \times\sum_l \binom{n-m}{l}\binom{m}{k-l}^{1/2}\binom{m}{k'-l}^{1/2}\ket{d_{k-l}^{m}} \bra{d_{k'-l}^{m}} ,
\end{eqnarray}
Finally, let us define $p=k-l$ and $q=k'-l$. Therefore, the state of $m$ qubits is given by
\begin{eqnarray}\nonumber
    \rho^{(m)}=\sum_{l=0}^{n-m}\sum_{p,q=0}^m a_{p+l} a_{q+l}^* \binom{n-m}{l}\binom{n}{p+l}^{-1/2}\times\\\times\binom{n}{q+l}^{-1/2}  \binom{m}{p}^{1/2}\binom{m}{q}^{1/2}\ket{d_{p}^{m}} \bra{d_{q}^{m}}.
\end{eqnarray}
For $m=1$ the above formula simplifies to 

\begin{eqnarray}\nonumber
    \rho^{(1)}=\tilde{\rho}=\sum_{l=0}^{n-1}\sum_{p,q=0}^{1} a_{p+l} a_{q+l}^* M_{pq}(n,l)\ket{p} \bra{q},
\end{eqnarray}
where
\begin{eqnarray}\nonumber
    M_{00}(n,l)=\frac{1}{n}(n-l),
\end{eqnarray}
\begin{eqnarray}\nonumber
    M_{01}(n,l)=M_{10}(n,l)=\frac{1}{n}\sqrt{(n-l)(l+1)}
\end{eqnarray}
and
\begin{eqnarray}\nonumber
    M_{11}(n,l)=\frac{1}{n}(l+1).
\end{eqnarray}

\subsection*{Calculation of initial distance}
In this section, we prove Eqs.(19) in main text.
Without losing generality, let initial state of a single qubit be
\begin{equation}
\tilde{\rho} = \left(\begin{array}{ccc}1&0\\0&0 \end{array} \right).
\end{equation}
The perturbed state is given by 
\begin{equation}
\tilde{\rho}_{\phantom{0}}\myprime = R_{\varphi}\,\tilde{\rho}\,R_{\varphi}^+
\end{equation}
where $R_{\varphi}=e^{i\frac{\varphi}{2}{\mathbf{m}}\cdot {\mathbf{s}}}$ denotes a single-qubit rotation about $\mathbf{m}$-axis,  with unit vector $\mathbf{m}=(m_x,m_y,0)$,  and ${\mathbf{s}}=(\sigma_x,\sigma_y,\sigma_z)$. The rotation angle $\varphi$ is assumed to be small ($\varphi \ll 1$). Linear in $\varphi$ approximation gives
\begin{equation}
\tilde{\rho}_{\phantom{0}}\myprime = \tilde{\rho}+i \frac{\varphi}{2} [{\mathbf{m}}\cdot {\mathbf{s}},\tilde{\rho}].
\end{equation}
Note that
\begin{equation}
[{\mathbf{m}}\cdot {\mathbf{s}},\tilde{\rho}]=i{\mathbf{m}}'\cdot {\mathbf{s}}
\end{equation}
where $\mathbf{m}'=(m_y,-m_x,0)$.
Now we have from Eq. (9) in the main text
\begin{equation}
D(\rho,\rho_{\phantom{0}}\myprime) = \frac{n}{2}{\text Tr}|\tilde{\rho}-\tilde{\rho}_{\phantom{0}}\myprime|=\frac{n\varphi}{4}{\text Tr}|{\mathbf{m}}'\cdot {\mathbf{s}}|=\frac{n\varphi}{2}.
\end{equation}


\subsection*{Classical kicked top}

The dynamics of the classical kicked top is given by the following equation
\begin{equation}
    \begin{pmatrix}
    x_{t+1}\\y_{t+1}\\z_{t+1}
    \end{pmatrix} = \begin{pmatrix}
        0 && \sin(\alpha x_t) && \cos(\alpha x_t)\\
        0 && \cos(\alpha x_t) && -\sin(\alpha x_t)\\
        -1 && 0 && 0
    \end{pmatrix}\begin{pmatrix}
    x_{t}\\y_{t}\\z_{t}
    \end{pmatrix}.
\end{equation}
The above equation was used to produce the right part of Fig.3 in the main text. In this figure, we use coordinates $(\phi,z)$ where $\phi=\text{arg}(x+iy)$.


\end{document}